\documentclass[referee]{aa} % for a referee version
\usepackage{graphicx}
\usepackage{natbib}
\usepackage{amssymb}
\usepackage{hyperref}
\usepackage{txfonts}
\usepackage{rotating,bm}
\usepackage{amsmath}
\usepackage{color}

\begin{document}
\title{A method to deconvolve mass ratio distribution from binary stars}
%\subtitle{Application to main--sequence field Stars}
\titlerunning{deconvolve mass distribution of binary stars}
\author{
Michel Cur\'{e} \inst{1}
\and
Diego F. Rial \inst{2}
\and
Julia Cassetti \inst{3}
\and
Alejandra Christen \inst{4}
\and 
Henri M.J. Boffin \inst{5}
}
\institute{
Instituto de F\'{i}sica y Astronom\'{i}a, Universidad de Valpara\'{i}so, Chile
\email{michel.cure@uv.cl}
\and
Departamento de Matem\'{a}ticas, Facultad de Ciencias Exactas y Naturales, 
Universidad de Buenos Aires and Instituto de Matem\'{a}tica Luis Santalo, IMAS–CONICET, Argentina\email{drial@mate.uba.ar}
\and
Universidad Nacional de General Sarmiento, Buenos Aires, Argentina, \email{jcassett@ungs.edu.ar}
\and
Instituto de Estad\'{i}stica, Pontificia Universidad Cat\'{o}lica de Valpara\'{i}so, Chile, \email{alejandra.christen@ucv.cl}
\and
ESO, Vitacura, Santiago, Chile. 
\email{hboffin@eso.org}
}
\date{Received ; Accepted }
\abstract{} {To better understand the evolution of stars in binary systems as well as to constrain the formation of binary stars, it is important to know the binary mass-ratio distribution. However, in most cases, i.e. for single-lined spectroscopic binaries, the mass ratio cannot be measured directly but only derived as the convolution of a function that depends on the mass ratio and the unknown inclination angle of the orbit on the plane of the sky.} 
{We extend our previous method 
to deconvolve this inverse problem (Cur\'e et al. 2014), i.e., we obtain as an integral the cumulative
 distribution function (CDF) for the mass ratio distribution.} {After a suitable transformation of variables it turns out that this problem is the same as the one for rotational velocities $v \sin i$, allowing a close analytic formulation for the CDF.  We then apply our method to two real datasets: a sample of Am stars binary systems, and a sample of massive spectroscopic binaries in the Cyg OB2 Association.} {We are able to reproduce the previous results of Boffin (2010) for the sample of Am stars, while we show that the mass ratio distribution of massive stars shows an excess of small mass ratio systems, contrarily to what was claimed by Kobulnicky et al. (2014). Our method proves very robust and deconvolves the distribution from a sample in just a single step.}
\keywords{
methods: analytical -- methods: data analysis -- methods: numerical -- methods: statistical -- 
stars: fundamental--parameters -- stars: binary
}
\maketitle

\section{Introduction\label{sec:intro}}

The knowledge of the mass ratio distribution of binary systems is crucial to 
understand how stars form and disentangle between the various proposed scenarios. Different ranges of mass ratios and 
separations have been obtained depending on the formation process (e.g., Halbwachs 1987; Maze \& Goldberg 1992; Clarke 2007, Kouwenhoven et al. 2009, Bate 2012).
Similarly, the mass ratio distribution can provide useful information on the formation channels of various classes of binary stars, as it impacts the evolution of a binary system (e.g., Boffin, Paulus \& Cerf 1992, Halbwachs et al. 2003).  
Unfortunately, in most cases, it is not possible to measure directly 
the mass ratio, as we have only access to the projected quantity $q \sin^3 i$, where $i$ is the inclination angle of the orbit on the plane of the sky and $q$ a function of the mass ratio (see below).

In order to deconvolve the mass ratio distribution function, it is
assumed that the inclination angles $i$ are uniformly distributed over the sphere. Based on this assumption, Heacox (1995) discussed different disentangling algorithms. Among them we find i) the {\it{analytic inverse}} that corresponds to the pioneer work of  Chandrasekhar \&  M\"unch (1950), giving a formal solution proportional to a derivative of an Abel's integral and ii) the {\it{iterative inverse}} that corresponds to the work of Cerf \& Boffin (1994; see Boffin 2010, 2012 for more recent applications of this method), using a Bayesian iterative method proposed by Lucy (1974). This method has 
the disadvantage that it possesses no convergence criteria (Bi \& Boerner, 1992) and the requested number of iterations has to be estimated carefully based on the signal to noise of the data, making sure not to obtain spurious peaks (Brown 2011, Boffin 2012). Despite these facts,  
Lucy's method is widely used in the astronomical community to disentangle distribution functions from different observations samples (Lucy 1994).

For the similar problem of rotational velocities, where the observed quantity is $v \sin i$, Cur\'e et al. (2014, hereafter paper I) enhanced the  work of  Chandrasekhar \& M\"unch (1950),  
integrating the probability distribution function (PDF)  and obtain the cumulative distribution function (CDF) for the velocity distribution, as a smooth function and in just one step, without iteration. %A main advantage of this method is that the CDF is attained in just one step, without the need of a convergence criteria usually necessary in iterative methods.

Following paper I, we apply this methodology to deconvolve the mass ratio
distribution from a sample of $q \sin^3 i$ under the usual assumption of uniform distribution of inclination angles. 
This article is structured as follows: in Section 2  we present a short summary of the 
work from paper I, and present the mathematical description of this binary mass ratio distribution problem. Section 3 is devoted to the calculation of the deconvolution of the distribution and to a discussion of the robustness of this method. In section 4 we  calculate the CDF from two real samples of spectroscopic binaries. In the last section we present our conclusions and future work.

\section{The Method\label{sec:method}}

In paper I we obtained the CDF of the rotational velocities of stars. We start from 
the integral representation of the problem,
\begin{equation}
\label{eq1}
f_{\tilde{Y}}(\tilde{y})=\int p(\tilde{y}\,|\,\tilde{x})\,f_{\tilde{X}}(\tilde{x})d\tilde{x}\, ,
\end{equation}
here  $\tilde{x}=v$ (rotational speed), $\tilde{y}=\tilde{x}\sin i$, being $i$ the inclination angle. 
This integral corresponds to a Fredholm integral of the first kind (Lucy 1994). Here $f_{\tilde{X}}$ is 
the function of interest (PDF) and the kernel $p$ of this integral is calculated assuming that the rotational axes are uniformly (randomly) distributed over the sphere. Replacing the kernel $p$ (see paper I), the Fredholm integral reads:
\begin{equation}
\label{eq-problema}
f_{\tilde{Y}}(\tilde{y})=\int_{\tilde{y}}^{\infty}\dfrac{\tilde{y}}{\tilde{x}\sqrt{\tilde{x}^{2}-\tilde{y}^{2}}}\,f_{\tilde{X}}(\tilde{x})d\tilde{x}.
\end{equation}
Chandrasekhar \& M\"unch (1950) deconvolved this equation to obtain the PDF:
\begin{equation}
\label{eq-f-metodo}
f_{\tilde{X}}(\tilde{x})=-\dfrac{2}{\pi}\tilde{x}^{2}\dfrac{\partial}{\partial \tilde{x}}\tilde{x}\int_{\tilde{x}}^{\infty}\dfrac{1}{\tilde{y}^{2}\sqrt{\tilde{y}^{2}-\tilde{x}^{2}}}\,f_{\tilde{Y}}(\tilde{y})d\tilde{y}.
\end{equation}

After integration of the PDF, the corresponding CDF reads:
\begin{equation}
\label{eq-F-metodo}
F_{\tilde{X}}(\tilde{x})=\,1-\dfrac{2}{\pi}\,\int_{\tilde{x}}^{\infty}
\left(\frac{\tilde{x}}{\sqrt{\tilde{y}^2-\tilde{x}^2}}+\arccos(\tilde{x}/\tilde{y})\right) f_{\tilde{Y}}(\tilde{y})\,d\tilde{y}\,  ,
\end{equation}
see paper I for details.

In this work we focus on a similar problem in astrophysics; the mass-ratio 
distribution of binary stars (see, e.g., Heacox 1995). For single-lined spectroscopic binaries, we have only access to the spectroscopic mass  function, given by:
\begin{equation}
y_m=\frac{K^3 \,P\, (1-e^2)^{3/2}}{2 \pi G}, 
\end{equation}
where $K$, $P$ and $e$ are the semi-amplitude of the radial velocity, orbital period and eccentricity. This mass function can be rewritten as:
\begin{equation}
y_m=\frac{r^3}{(1+r)^2} m_p\; \sin^3i,
\end{equation}
where $m_p$ is the mass of the primary, $m_c$ the mass of the companion, $r=m_c/m_p$ is the mass ratio and $i$ corresponds 
to the inclination angle between the orbital plane and the plane of the sky.
Thus, the observed reduced spectroscopic mass function is: 
\begin{equation}
\label{y-sample}
y=y_m/m_p = q \, \sin^3i,
\end{equation}
where $q\equiv r^3/(1+r)^2$.

For this case, the kernel $p$, following Boffin, Cerf \& Paulus (1992) and Cerf \& Boffin (1994), reads:
\begin{equation}
p(y\,|\,q)=\frac{1}{3q^{1/3}y^{1/3}\sqrt{q^{2/3}-y^{2/3}}}.
\end{equation}

Cerf \& Boffin (1994) %where the fist to apply  Lucy's (1974) method for this case, they 
found that the Friedholm integral eq. (\ref{eq1}) can the be written as:
\begin{equation}
\label{pdfQ}
f_{Y}(y)=\int_{y}^{1/4}\frac{1}{3q^{1/3}y^{1/3}\sqrt{q^{2/3}-y^{2/3}}}f_{Q}(q)dq.
\end{equation}

Now we have to solve this inverse problem, i.e., obtain $f_{Q}(q)$ from Eq. (\ref{pdfQ}).

\section{Deconvolving the mass ratio distribution} 
Defining $Y=x^{3}$, then $f_{X}(x)=3x^{2} \,f_{Y}(x^{3})$,  Eq.~(\ref{pdfQ}) becomes:
\begin{align*}
f_{X}(x)=\int_{x^{3}}^{1/4}\frac{x}{q^{1/3}\sqrt{q^{2/3}-x^{2}}}f_{Q}(q)dq,
\end{align*}
applying  the following transformation of variables $q=p^{3}$, Eq.~(\ref{pdfQ}) now reads:
%\begin{eqnarray}
\begin{equation}
\label{pdfVSini}
f_{X}(x)=\int_{x}^{4^{-1/3}}\frac{x}{p\sqrt{p^{2}-x^{2}}}f_{P}(p)dp\, .\\
\end{equation}
%here we used $f_{Q}(q)\,dq = 3 p^2 f_{P}(p)\,dp$.

This last expression is the same as Eq.~(\ref{eq-problema}),  
which has the following solution:
\begin{equation}
\label{Solution-CDF}
F_{P}(p)=1-\frac{2}{\pi}\int_{p}^{4^{-1/3}}
\left(\frac{p}{\sqrt{x^{2}-p^{2}}}+\arccos(p/x)\right)
f_{X}(x)dx
\end{equation}
In order to avoid numerical instabilities in the integration process of Eq. (\ref{Solution-CDF}), 
we propose to use the following variable, $z=\sqrt{x^{2}-p^{2}}$, or equivalently 
$x=\sqrt{p^{2}+z^{2}}$ and $dx=z\,dz/\sqrt{p^{2}+z^{2}}$.
Therefore,  Eq. (\ref{Solution-CDF}) transforms to:
\begin{equation}
\label{Solution-CDF-s}
F_{P}(p)=1-\frac{2}{\pi}\int_{0}^{\sqrt{4^{-2/3}-p^{2}}}
K(p,z)
f_{X}\left(\sqrt{p^{2}+z^{2}}\right)dz
\end{equation}
where the function $K$ is defined as:
\begin{equation}
K(p,z)=\frac{p}{\sqrt{p^{2}+z^{2}}}+\frac{z}{\sqrt{p^{2}+z^{2}}}\arccos\left(\frac{p}{\sqrt{p^{2}+z^{2}}}\right)
\end{equation}
Finally. the CDF as function of the original variable ($r$), reads:
\begin{equation}
F_R(r)=F_P\left(\frac{r}{(1+r)^{2/3}}\right)
\end{equation}
\subsection{Logarithmic Variables}

Boffin (2010) has shown that when dealing with  spectroscopic mass function distributions, which span 
 a wide range of values, it is often necessary to use logarithmic variables. Let's define therefore $z=e^{-s}$, 
 such that the previous expression (Eq. \ref{Solution-CDF-s}) is written as:
\begin{equation}
\label{Solution-CDF-log}
F_{P}(p)=1-\frac{2}{\pi}\int_{s_{0}}^{\infty}K(p,e^{-s})
f_{X}\left(\sqrt{p^{2}+e^{-2s}}\right)e^{-s}ds
\end{equation}
where $s_{0}=-\ln(4^{-2/3}-p^{2})/2$

\section{CDF Application to Real Data-Sets}
In this section we will apply the proposed method for two samples of real astronomical data to obtain the CDF distribution as function of the mass ratio $r$. In addition we will compare our results with the Lucy's (1974) method.
\subsection{Real Data-Set 1: Am binaries}
Boffin (2010) derived the mass ratio distribution for a sample of 134 Am binaries using 
Lucy's (1974) method, as implemented by Boffin et al. (1993). He found that this distribution can be approximated by a uniform distribution. 

Smalley et al. (2014) reported the discovery of 70 eclipsing Am systems using light curves from the SuperWASP project.  Based on these light curves they estimated the mass ratio distribution of Am stars. When taking into account the fact that low-mass companions to Am stars may be too small and too faint to be detected as eclipsing, they found a mass ratio distribution consistent with that of Boffin (2010).

\begin{figure}
\begin{center}
\includegraphics[scale=1.]{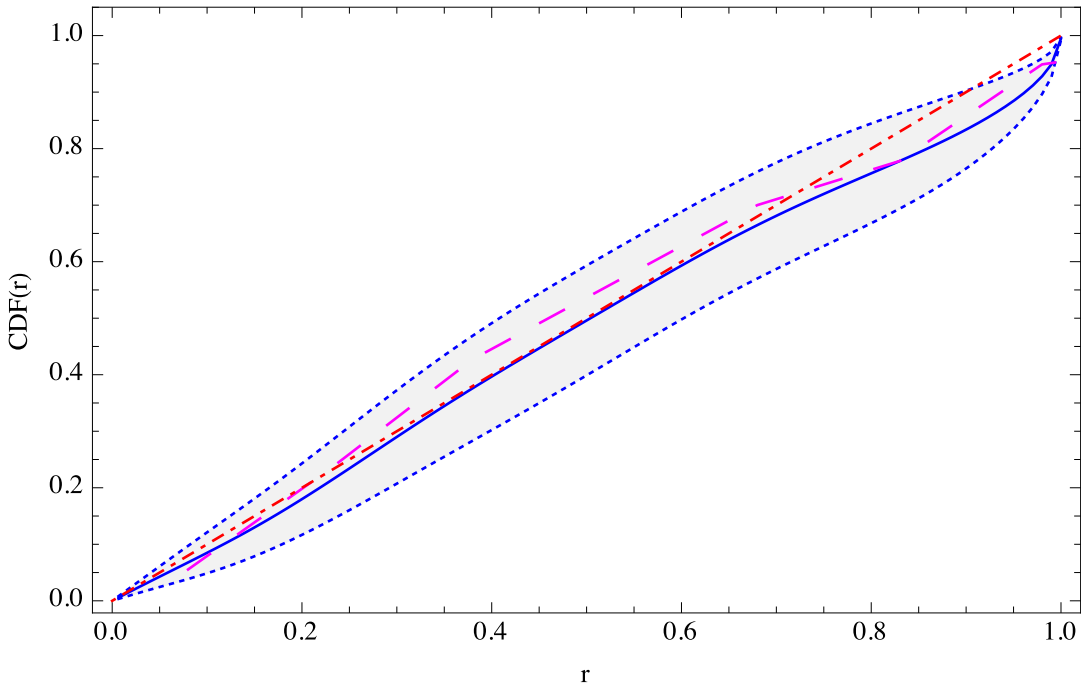} 
\caption{The estimated CDF using our method in solid line. In short--dashed lines the 95\% confidence interval is also shown. The CDF using  Lucy’s (1974) method from Boffin (2010) is shown as long--dashed line while the theoretical CDF from a uniform distribution is in dot--dashed line. See text for details. \label{fig1}}
\end{center}
\end{figure}

In order to constrain the range of $y = q \sin^3 i$ as function of $r$ we have to take in account the following:  $\sin i$ is smaller than one, then we are left with $q=r^3/(1+r)^2$. Similarly, for physical reasons, the mass ratio has to be smaller or equal to 1 ($r\leq1$), otherwise if the companion is more massive, as they are main sequence stars, it would be more luminous, and the system would not appear as it is. Thus $q$ has to be smaller than $1/4$.\\
The reason why some values of $y$ appear larger than $0.25$ are most likely due to observational errors on the spectroscopic 
mass functions or a wrong estimate of the primary mass. As such cases are very few, we prefer remove them from the sample 
to have a more homogeneous one.\\
Considering these reasons, our method has been developed in the range: $0<y<1/4$. \\
We have filtered Boffin's (2010) data-set for values of 
$y$ (Eq. \ref{y-sample}) in that range getting 119 data values. Then we calculated the CDF as function of $r$, shown as solid line in Figure \ref{fig1}.

To calculate the error of our method, we use the bootstrap method (Efrom 1993). 
The bootstrap method attempts to determine the probability distribution from the data itself, without recourse to central limit theorem. This 
method is a way to estimate the error of the sample. Basic idea of  Bootstrap is in our case: Compute from a sample the CDF; create an artificial sample by randomly drawing elements from the original sample, some elements will be picked more than once; Then compute the new CDF and repeat $100$ - $1000$ times this procedure to look to the distribution of these CDFs.

Thus, we create $500$ bootstrap samples from the original sample from Boffin (2010) and calculated for each of these bootstrap--samples the corresponding CDF. To obtain a 95\% confidence interval we use the same procedure as in Cur\'e et al. (2014), i.e., calculate confidence interval with normal standard percentile $Z_{\alpha/2}$, namely: 
\begin{equation}
\widehat{CDF} \pm Z_{\alpha/2} \widehat{se}\, ,
\end{equation}
where $\widehat{se}$ is the standard deviation of all Bootstrap samples, and $Z_{\alpha/2}$ is the value in which the standard normal distribution accumulate $97.5\%$ of the area under its PDF and $\alpha=0.05 (=1-0.95)$ is the complement of the confidence.\\
These upper and lower confidence limits are shown in dashed line in  Figure \ref{fig1} and the area between both curves is light--gray shadowed. In addition, we plotted the result using Lucy's (1974) method from Boffin (2010).\\

Our method for the CDF is in complete agreement with the Lucy's method, i.e., arriving at the same conclusion that the sample 
is compatible with a uniform distribution. In addition to the advantage that our method does not need any iteration criteria and it 
deconvolves the mass ration CDF in one step, it can also provide (using the bootstrap method) confidence intervals for this CDF.

\subsection{Real Data-Set 2: Binaries in the Cyg OB2 Association \label{secCDF}}

\begin{figure}
\begin{center}
\includegraphics[scale=1.]{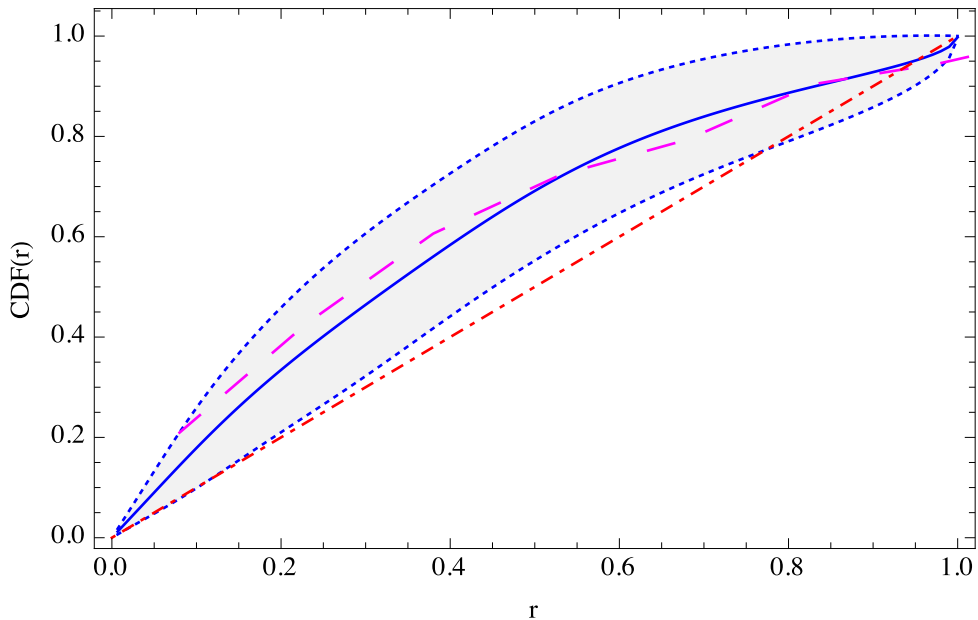} 
\caption{The estimated CDF for Cyg OB2 Association using the sample from Kobulnicky et al. (2014). Our method as solid lines with the confidence intervals as short--dashed lines. Lucy’s method as long--dashed line. The theoretical CDF from a uniform distribution is in dot--dashed line. \label{fig2}}
\end{center}
\end{figure}

Based on previous work spanning several years, Kobulnicky et al. (2014) present a sample of 48 massive 
multiple-star systems known in the Cygnus OB2 Association and analyzed their orbital properties. In particular, 
using a Monte Carlo method, they conclude that the observed distribution of mass ratios is consistent with a 
uniform distribution, even though they appreciate that due to incompleteness, the bins at $r<0.2$ are likely 
underestimated. 
We have used their data: spectroscopic mass function sample and adopted primary mass, assuming that all 
systems are single-lined spectroscopic binaries. After filtering in the range $0<y<1/4$ the sample consist of 47 data points. 
Then we compute the mass ratio distribution with the confidence intervals based on our methodology, 
as well as on the Lucy's method as implemented by Boffin (2010).  The results are shown in  Figure \ref{fig2}.\\

It is clear that instead of finding a uniform distribution over the full range of mass ratios, we find an excess of 
systems in the whole interval $0<r<1$ as compared with an uniform distribution. A comparison between both methods shows a similar behavior in almost the entire $r$--interval, but this does not correspond to an uniform distribution. Lucy's method-CDF lies inside the confidence interval of our CDF, therefore we conclude that statistically are the same CDF. On the other hand, Figure 3 shows  the CDF vs. $\log(r)$, and  
it seems that the CDF is linear in $\log(r)$ from $r\gtrsim 0.2$. This indicates that instead of a uniform distribution, the sample of massive stars is more consistent with an excess of low mass companions.

As mentioned above, the interval $r<0.2$ suffers from incompleteness, so the real contribution of these low-mass companions may be even larger than found here. This may have important consequences for the final binary fraction of massive stars and should be taken into account in further observational campaigns to aim at discovering companions to massive stars.

\begin{figure}
\begin{center}
\includegraphics[scale=1.]{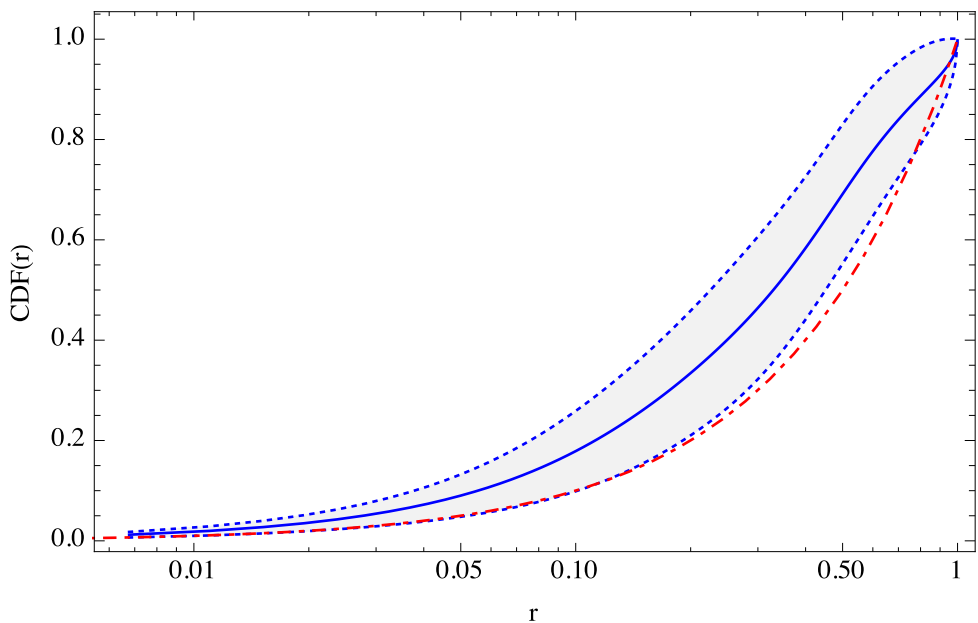} 
\caption{The estimated CDF plotted as function of $\log(r)$ as solid line with the confidence intervals as short--dashed lines. The theoretical CDF from a uniform distribution is in dot--dashed line. This estimated CDF seems to be linear in $\log(r)$ from $r\gtrsim 0.2$.\label{fig3}}
\end{center}
\end{figure}
%From the above calculations, we conclude that our method recovers the CDF with vey high confidence.

\section{Conclusion}
In this work we have obtained  the cumulative distribution function of ''de-projected'' mass ratios from binary systems. 
After a suitable change of variables, we showed that this problem is the same as the problem of the distribution of 
rotational velocities from stars, therefore its solution is known (paper I).

This method allows us to obtain the CDF without numerical instabilities caused by the use of derivative (Chandrasekar \& M\"unch 1950) in just one step without needing any convergence criteria as needed by the widely used iterative method of Lucy (1974). In addition, confidence 
intervals can be attained using bootstrap algorithm.\\

We have applied our method to two real samples of binaries. While we find the same result as Boffin (2010) 
for the sample of Am spectroscopic binaries, we find that the mass ratio distribution of massive binaries in the
Cyg2 OB Association is presenting an excess of systems with mass ratios $r< 0.2$, and is highly probable that
this sample does not comes from a uniform distribution, in contradiction with the analysis of Kobulnicky et al. 
(2014).

In the future, we plan to extend the applicability of our model in the case of a general function describing an 
arbitrary orientation of inclination angles (see, e.g., Boffin 2012 for reasons why this could be needed) and  
study these distributions (rotational velocities or mass ratio distributions) in a more precise formulation. 

\begin{acknowledgements}
MC thanks the support of FONDECYT project 1130173 and Centro de Astrof\'isica de Valpara\'iso. JC thanks the financial support from project: "Ecuaciones Diferenciales y An\'alisis Num\'erico" - Instituto de Ciencias, Instituto de Desarrollo Humano e Instituto de Industria - Universidad Nacional de General Sarmiento. DR acknowledge the support of project PIP11420090100165, CONICET. AC thanks the support from PUCV's projects: 126.711/2014 and 37.375/2014.
\end{acknowledgements}

\end{document}